\documentclass[preprint,amsmath,aps,10pt,superscriptaddress,letterpaper,tightenlines,showkeys]{revtex4}
\usepackage{bm}
\usepackage{stmaryrd}
\usepackage{amsmath}
\usepackage{amssymb}
\usepackage{graphicx}
\usepackage{textcomp}
\usepackage{calrsfs}
\usepackage{yfonts}
\usepackage{color}
\usepackage{amsthm}
\usepackage{lipsum}
\newcommand{\haf}{{\frac{1}{2}}}

\newcommand{\la}{{\langle}}
\newcommand{\ra}{{\rangle}}

\newcommand{\p}{{\partial}}

\begin{document}
\title{Dissipative noninteracting scalar field theory: A covariant formulation}

\author{A. Refaei}
\email{refaei@iausdj.ac.ir}
\affiliation{Department of Physics, Sanandaj branch, Islamic Azad University, Sanandaj, Iran.}

\author{F. Kheirandish}
\email{fkheirandish@yahoo.com}
\affiliation{Department of Physics,
Faculty of Science, The University of Isfahan,
Isfahan, Iran.}
\begin{abstract}
Caldeira-Leggett model of reservoir is generalized to a reservoir modeled by a continuum of real Klein-Gordon fields, instead of harmonic oscillators. A quantum Langevin type dissipative equation is obtained for the scalar field. The susceptibility of the medium is defined in terms of the reservoir Green's function and the coupling function satisfying causality condition. The connection between the coupling function and the susceptibility of the medium is found to be a Hankel transform from which the coupling function can be determined in terms of the susceptibility of the medium. Noise currents and their fluctuation-dissipation relation are obtained. In a homogeneous medium or reservoir, explicit form of the quantum scalar field, and its large-time limit, are found.
\end{abstract}
\keywords{Scalar field; noninteracting; Covariant; Dissipation; Quantization; Susceptibility function; Coupling function; Reservoir}
\maketitle
\section{Introduction}
Dissipative scalar field theories appear in a variety of problems in physics, for example in Casimir physics, a scalar field is the fluctuating field interacting linearly with some external matter fields defined inside or over some specific surfaces. In this case dissipation appears in the equation of motion of the scalar field trough the susceptibility of the medium. The scalar field or the main field, satisfies Dirichlet or Neumann boundary conditions on these surfaces causing fluctuating-induced forces among matter fields \cite{Casimir}. Another interesting example is the reheating of the universe after an inflationary epoch has passed \cite{Guth,Brand}.

A similar situation appears in the electromagnetic field quantization in the presence of metals or dielectrics where due to the specified geometry the electromagnetic field can be basically considered as two independent scalar fields. In this case one uses the Huttner-Barnett model \cite{Huttner,Kheir} to quantize the electromagnetic field in the presence of matter fields modelled by continuum sets of harmonic oscillators modeling electric and magnetic properties of the medium. For a covariant approach to quantize the electromagnetic field in the presence of matter fields see \cite{Amoo}.

In 1983, Caldeira and Leggett proposed a model to incorporate dissipation into the equation of motion based on a microscopic theory \cite{Caldeira}. They modeled the matter field or reservoir as a combination of non interacting harmonic oscillators with different mass and frequencies which were coupled to the main system linearly, knowing as system-reservoir model. In this model, by writing Heisenberg equations for both system and reservoir degrees of freedom and eliminating the reservoir degrees of freedom one finds a Langevin type equation for the main system including a noise field naturally induced from the vacuum fluctuations of the matter or reservoir system. This method has been applied to a variety of problems for example to the polaron motion in a crystal or the dynamics of Josephson junction \cite{Weiss} with successful experimental predictions \cite{Voss,Martinis}. In Caldeira-Legget and Huttner-Barnett model of the matter fields, the Lagrangians associated with these matter fields are not Lorentz-Invariant and so the total Lagrangian density is not lorentz invariant.

In the present article, inspired by Caldeira-Leggett model, we present a covariant formulation of a non-interacting dissipative quantum scalar field by modeling the medium or reservoir as a continuum of Klein-Gordon fields instead of harmonic oscillators and find a quantum Langevin type dissipative equation for the scalar field. We follow a canonical approach to quantize the field-reservoir model though a path integral quantization of the system is also possible and may be more effective while considering interacting quantum field theories. The susceptibility of the medium is defined in terms of the reservoir Green function and coupling function satisfying causality condition. The connection between the coupling function and the susceptibility of the medium is found to be a Hankel transform. In the special case where the reservoir is assumed to be homogeneous, the explicit form of the quantum scalar field is obtained. Knowing the explicit form of the quantum field, one can proceed and construct Fock's space or find important quantities like the two-point function or Feynman Green function. Including finite temperature corrections is also straightforward since the field is expressed in terms of the creation and annihilation operators of the medium which can be easily thermalized. A more realistic problem appears when one considers an interacting field theory where a path integral approach seems to be more effective, at least from perturbative point of view.
\section{Lagrangian}
 Throughout the article, without losing any generality, we assume that the fields are defined in $1+1$-dimensional space-time and use natural units $\hbar=c=1$, for notational convenience. For a real scalar field $\varphi(x,t)$, interacting with an environment defined by real scalar fields $Y_\omega (x,t)$, we assume the following covariant Lagrangian density for the field-environment system 
\begin{eqnarray}\label{L}
  \mathcal{L} &=& \haf\,\p_\mu\,\varphi (x,t)\,\p^\mu\,\varphi (x,t)-\haf\,m^2 \varphi^2 (x,t)\nonumber\\
  && + \,\haf\,\int_0^{\infty}d\omega\,[\p_\mu\,Y_\omega (x,t) \,\p^\mu\,Y_\omega (x,t)-\omega^2 Y^2_\omega (x,t)]\nonumber\\
  && +\int_0^{\infty}d\omega\,f(\omega)\,\varphi(x,t)\,Y_\omega(x,t),
\end{eqnarray}
where $f(\omega)$ is a coupling function which is assumed to be homogeneous here, that is independent on space-time coordinates. In a nonhomogeneous medium it depends on space-time variables and should be considered as a scalar classical field to keep the total Lagrangian covariant. The difference between the Lagrangian density in Eq.(\ref{L}) and in previous models \cite{Huttner,Kheir}, is the modified matter Lagrangian density, which is now assumed to be a continuum of $1+1$-dimensional real Klein-Gordon fields instead of a continuum of harmonic fields. Note that the index of continuity is the frequency parameter $\omega\in(0,\infty)$. From Euler-Lagrange equations, we find the classical equations of motion for the fields as
\begin{equation}\label{Efi}
(\partial^2+m^2)\,\varphi(x,t)=\int_0^\infty d\omega\,f(\omega)\,Y_\omega(x,t),
\end{equation}
and
\begin{equation}\label{EY}
  (\partial^2+\omega^2)\,Y_\omega(x,t)=f(\omega)\,\varphi(x,t),
\end{equation}
where $\partial^2=\p^2_t-\p^2_x$. These are coupled integro-differential equations. In the next section, we quantize the model via canonical quantization approach and find similar coupled integro-differential equations for the quantum fields which we are interested in.
\section{Canonical quantization}
From the Lagrangian density (\ref{L}), the conjugate momenta corresponding to the fields $\varphi$ and $Y_\omega$ are defined by
\begin{equation}\label{momfi}
  \pi (x,t) = \frac{\p \mathcal{L}}{\p \dot{\varphi}}=\dot{\varphi}(x,t),
\end{equation}
\begin{equation}\label{momYi}
  \Pi_\omega (x,t) = \frac{\delta \mathcal{L}}{\delta \dot{Y}_\omega}=\dot{Y}_\omega (x,t).
\end{equation}
To quantize the theory canonically, the following equal-time commutation relations are be imposed on the fields and their conjugate momenta
\begin{equation}\label{comfi}
 [\hat{\varphi}(x,t),\,\hat{\pi}(x',t)]=i\,\delta(x-x'),
 \end{equation}
 \begin{equation}\label{comYi}
 [\hat{Y}_\omega (x,t),\,\hat{\Pi}_{\omega'} (x',t)=i\,\delta(\omega-\omega')\delta(x-x'),
\end{equation}
and the other commutation relations are identically zero. Using the canonical momenta Eqs.(\ref{momfi},\ref{momYi}) and the Lagrangian density Eq.(\ref{L}), we find Hamiltonian density from the definition
\begin{equation}
  \hat{\mathcal{H}} = \hat{\pi}\dot{\hat{\varphi}}+\int_0^\infty d\omega\,\hat{\Pi}_\omega \dot{\hat{Y}}_\omega-\hat{\mathcal{L}}.
\end{equation}
The Hamiltonian of the field-environment system is defined by
\begin{eqnarray}\label{H}
  \hat{H} &=& \int\limits_{-\infty}^{\infty} dx\,\hat{\mathcal{H}},\nonumber\\
  &=& \int\limits_{-\infty}^{\infty} dx\, \bigg\{\haf (\hat{\pi}^2+(\p_x\hat{\varphi})^2+m^2\hat{\varphi}^2)\nonumber\\
   && + \,\haf\int_0^\infty d\omega\,(\Pi^2_\omega+(\p_x \hat{Y}_\omega)^2+\omega^2 \hat{Y}^2_\omega) \nonumber\\
   && - \int_0^\infty d\omega\,f(\omega)\,\hat{\varphi}(x,t)\,\hat{Y}_\omega (x,t)\bigg\}.
\end{eqnarray}
Having the Hamiltonian and using the Heisenberg equations
\begin{eqnarray}
  i\,\frac{\p \hat{\varphi}(x,t)}{\p t} &=& [\hat{\varphi}(x,t),\hat{H}], \\
  i\,\frac{\p \hat{Y}_\omega(x,t)}{\p t} &=& [\hat{Y}_\omega(x,t),\hat{H}],
\end{eqnarray}
we obtain the equations of motion for the field $\hat{\varphi}$ and matter fields $\hat{Y}_\omega$, respectively as
\begin{equation}\label{Qfi}
(\partial^2+m^2)\,\hat{\varphi}(x,t)=\int_0^\infty d\omega\,f(\omega)\,\hat{Y}_\omega(x,t),
\end{equation}
\begin{equation}\label{QY}
(\partial^2+\omega^2)\,\hat{Y}_\omega(x,t)=f(\omega)\,\hat{\varphi}(x,t),
\end{equation}
which are similar to their classical counterparts Eqs.(\ref{Efi},\ref{EY}). The formal solution of Eq.(\ref{QY}) is
\begin{equation}\label{SY}
  \hat{Y}_\omega(x,t)=\hat{Y}^N_\omega(x,t)-\int dx'\,G_\omega(x-x',t-t')\,f(\omega)\,\hat{\varphi}(x',t'),
\end{equation}
where $G_\omega (x-x',t-t')$ is the Green's function of Eq.(\ref{QY}) satisfying
\begin{equation}\label{EG}
 (\partial^2+\omega^2)\,G_\omega(x-x',t-t')=-\delta(x-x')\delta(t-t').
\end{equation}
To find the explicit form of the Green's function $G_\omega (x-x',t-t')$, we first take the temporal Laplace transform of both sides of Eq.(\ref{EG}), putting the initial conditions equal to zero
\begin{equation}\label{LG}
\frac{d^2}{dx^2}\,\tilde{G}_\omega (x-x',s)-(s^2+\omega^2)\,\tilde{G}_\omega (x-x',s)= \delta(x-x'),
\end{equation}
the solution of Eq.(\ref{LG}) is 
\begin{equation}\label{SLG}
  \tilde{G}_\omega (x-x',s)=-\frac{1}{2\sqrt{s^2+\omega^2}}\,e^{-\sqrt{s^2+\omega^2}\,|x-x'|}.
\end{equation}
Now taking the inverse Laplace transform we find \cite{Gradshteyn}
\begin{equation}\label{Green}
  G_\omega (x-x',t-t')=-\haf\,\theta(t-t'-|x-x'|)\,J_0 (\omega\sqrt{(t-t')^2-|x-x'|^2}\,),
\end{equation}
where $\theta(x)$, is Heaviside step function and $J_0 (x)$, is Bessel function of the first kind and zero order.

The homogeneous solution $\hat{Y}^N_\omega (x,t)$ satisfies the free space Klein-Gordon equation
\begin{equation}\label{EYN}
 (\partial^2+\omega^2)\,\hat{Y}^N_\omega (x,t)=0,
\end{equation}
and can be interpreted as the noise fields or quantum vacuum fluctuating fields. The noise fields can be expanded in terms of the eigen-modes of the Klein-Gordon equation as
\begin{eqnarray}\label{SYN}
 \hat{Y}^N_\omega (x,t) &=& \int \frac{dk}{\sqrt{2\pi (2\omega_k )}}\,\bigg[\hat{a}_k (\omega)\,e^{i (kx-\omega_k t)}
 +\hat{a}^{\dag}_k (\omega)\,e^{-i (kx-\omega_k t)} \bigg],
\end{eqnarray}
where $\omega_k=k_0=\sqrt{k^2+\omega^2}$, and the creation $(\hat{a}^\dag_k)$ and annihilation $(\hat{a}_k)$ operators satisfy the usual commutation relations induced from the canonical commutation relation Eq.(\ref{comYi})
\begin{equation}
[\hat{a}_k (\omega),\,\hat{a}^\dag_{k'} (\omega')]=\delta(\omega-\omega')\delta(k-k').
\end{equation}
From Eqs.(\ref{SY},\ref{SYN}), we can rewrite Eq.(\ref{Qfi}) as
\begin{equation}\label{Lanfi}
  (\partial^2+m^2)\,\hat{\varphi}(x,t)+\int\int dx'\,dt'\,\gamma(x-x',t-t')\,\hat{\varphi}(x',t')=\hat{J}^N(x,t),
\end{equation}
where the memory function $\gamma(x,t)$ or the susceptibility of the medium is defined by
\begin{equation}\label{mem1}
 \gamma(x-x',t-t')=\int_0^\infty d\omega\,f^2(\omega)\,G_\omega(x-x',t-t'),
\end{equation}
and the noise current $\hat{J}^N (x,t)$ is given by
\begin{equation}\label{noise}
 \hat{J}^N(x,t)=\int_0^\infty d\omega\,f(\omega)\,\hat{Y}^N_\omega(x,t).
\end{equation}
From Eq.(\ref{noise}), we find the commutation relation
\begin{equation}\label{fluc}
 [\hat{J}^N(x,t),\,\hat{J}^N(x',t')]=i\,[\theta(t-t')\,\gamma(x-x',t-t')-\theta(t'-t)\,\gamma(x'-x,t'-t)].
\end{equation}
Using Eq.(\ref{fluc}), one can easily show that
\begin{equation}\label{fluc-diss}
  \la 0|\hat{J}^N_{+}(x,\omega)\,\hat{J}^N_{-}(x,\omega')|0\ra=4\pi\delta(\omega-\omega')\mbox{Im}[\tilde{\gamma}(x-x',\omega)],
\end{equation}
known as the fluctuation-dissipation theorem. In Eq.(\ref{fluc-diss}), $\hat{J}^N_{+}(x,\omega)$ and $\hat{J}^N_{-}(x,\omega)$, are the Fourier transform of the positive and negative frequency parts of the field $\hat{J}^N(x,t)$ defined by Eqs.(\ref{SYN},\ref{noise}).
By inserting Eq.(\ref{Green}) for the Green's function into Eq.(\ref{mem1}) we find
\begin{equation}\label{mem2}
  \gamma(x-x',t-t')=-\haf\,\theta(t-t'-|x-x'|)\int_0^\infty d\omega\,f^2(\omega)\,J_0 (\omega\sqrt{(t-t')^2-|x-x'|^2}\,).
\end{equation}
Eq.(\ref{mem2}) is a Hankel transform and can be inverted to find the coupling function in terms of the memory function as
\begin{equation}\label{coupling}
  f^2 (\omega)=-\omega\,\int_0^\infty du\,u\,\gamma(u)\,J_0 (\omega u).
\end{equation}
The left hand side of Eq.(\ref{coupling}) is positive and this condition will impose a physical constraint on acceptable memory functions, that is the right hand side of Eq.(\ref{coupling}) should be positive. In non-relativistic models the constraint was the positivity of the imaginary part of the memory function or the susceptibility, in order to have dissipation in the system. Negativity of the imaginary part of the susceptibility leads to gain in the system and we do not discuss it here.

Taking the Fourier-Laplace transform of both sides of Eq.(\ref{Lanfi}) with respect to spatial-temporal dependence respectively, leads to
\begin{eqnarray}\label{fiks}
  \tilde{\hat{\varphi}}(k,s) &=& \frac{s}{s^2+k^2+m^2+\tilde{\gamma}(k,s)}\,\hat{\varphi}(k,0)+\frac{1}{s^2+k^2+m^2+\tilde{\gamma}(k,s)}\,\dot{\hat{\varphi}}(k,0)\nonumber\\
  &+& \frac{\tilde{\hat{J}}^N (k,s)}{s^2+k^2+m^2+\tilde{\gamma}(k,s)},
\end{eqnarray}
where
\begin{equation}\label{gks}
  \tilde{\gamma}(k,s)=-\int_0^\infty d\omega\,\frac{f^2(\omega)}{s^2+\omega^2+k^2}.
\end{equation}
In deriving Eq.(\ref{gks}) we used
\begin{equation}\label{}
  \tilde{G}_\omega(k,s)=-\frac{1}{s^2+\omega^2+k^2},
\end{equation}
which can be easily obtained from Eq.(\ref{LG}) using spatial Fourier transform. The Fourier-Laplace transform of the  noise field is given by
\begin{equation}\label{}
  \tilde{\hat{J}}^N (k,s)=\int_0^\infty d\omega\,f(\omega)\,\frac{2\pi}{\sqrt{4\pi \omega_k}}\,\bigg\{\frac{1}{s+i\omega_k}\,\hat{a}_k (\omega)+
  \frac{1}{s-i\omega_k}\,\hat{a}^\dag_k (\omega)\bigg\}.
\end{equation}
For later convenience, let us define the functions $\beta(k,t)$ and $\alpha(k,t)$ by
\begin{equation}\label{beta}
  \beta(k,t)=L^{-1}\bigg[\frac{s}{s^2+k^2+m^2+\tilde{\gamma}(k,s)}\bigg],
\end{equation}
\begin{equation}\label{alfa}
  \alpha(k,t)=L^{-1}\bigg[\frac{1}{s^2+k^2+m^2+\tilde{\gamma}(k,s)}\bigg]=\int_0^t dt'\,\beta(k,t'),
\end{equation}
where $L^{-1}$ is the inverse laplace transform operator. From Eq.(\ref{fiks}) and definitions Eqs.(\ref{beta},\ref{alfa}), we have
\begin{eqnarray}\label{fikt}
\tilde{\hat{\varphi}}(k,t) &=& \beta(k,t)\,\tilde{\hat{\varphi}}(k,0)+\alpha(k,t)\,\dot{\tilde{\hat{\varphi}}}(k,0)\nonumber\\
&& + \int_0^t dt'\,\alpha(k,t')\int_0^\infty d\omega\,f(\omega)\sqrt{\frac{\pi}{\omega_k}}\,\bigg(\hat{a}_k(\omega)e^{-i\omega_k (t-t')}+
\hat{a}^\dag_{-k}(\omega)e^{i\omega_k (t-t')}\bigg).
\end{eqnarray}
Therefore, the explicit form of the field $\hat{\varphi}(x,t)$ is given by
\begin{eqnarray}\label{fi}
  \hat{\varphi}(x,t) &=& \int_{-\infty}^{\infty}\frac{dk}{2\pi}\,e^{ikx}\,\beta(k,t)\,\tilde{\hat{\varphi}}(k,0)+\int_{-\infty}^{\infty}\frac{dk}{2\pi}\,e^{ikx}\,\alpha(k,t)\,
  \dot{\tilde{\hat{\varphi}}}(k,0)\nonumber\\
  && + \int_0^t dt'\,\int_0^\infty d\omega\,f(\omega)\,\int_{-\infty}^{\infty} \frac{dk}{\sqrt{2\pi\,2\omega_k}}\,\alpha(k,t')\,\bigg\{\hat{a}_k(\omega)e^{i[kx-\omega_k (t-t')]}+
\hat{a}^\dag_{k}(\omega)e^{-i[kx-\omega_k (t-t')]}\bigg\}.
\end{eqnarray}
At the large-time limit $(t\gg1/\omega_k)$, we have
\begin{eqnarray}\label{alfalim}
\int_0^{t\gg1/\omega_k} dt'\,\alpha(k,t')\,e^{i\omega_k t'} &\approx &\tilde{\alpha}(k,s=-i\omega_k),\nonumber\\
&& =\frac{1}{m^2-\omega^2+\tilde{\gamma}(k,-i\omega_k)},
\end{eqnarray}
where we made use of Eq.(\ref{alfa}). Also from Eq.(\ref{gks}), we have
\begin{equation}\label{gg}
 \tilde{\gamma}(k,-i\omega_k)=-\int_0^\infty d\omega'\,\frac{f^2(\omega')}{\omega'^2-\omega^2}.
\end{equation}
Note that due to the dissipation, the dominant term in Eq.(\ref{fi}) is the last term, this term could also be obtained directly from Fourier transform instead of taking Laplace transform but in deriving Eq.(\ref{fi}) we were interested in the derivation of the complete solution containing the homogeneous part in addition to the particular solution. Now, using Eq.(\ref{alfalim}) we find the large-time limit of the field $(\hat{\varphi}_{c}(x,t))$ as
\begin{eqnarray}
  \hat{\varphi}_{c}(x,t)=
  \int_{-\infty}^{\infty} \frac{dk}{\sqrt{2\pi\,2\omega_k}}\, \int_0^\infty d\omega\,\frac{f(\omega)}{m^2-\omega^2+\tilde{\gamma}(k,-i\omega_k)}\,
  \bigg\{\hat{a}_k(\omega)e^{i[kx-\omega_k t]}+
\hat{a}^\dag_{k}(\omega)e^{-i[kx-\omega_k t]}\bigg\}.
\end{eqnarray}
Knowing the explicit form of the quantum field, one can proceed and construct Fock's space or find important quantities like the two-point function or Feynman Green function. Including finite temperature corrections is also straightforward since the field is expressed in terms of the creation and annihilation operators of the medium which can be easily thermalized. A more realistic problem appears when one considers an interacting field theory where a path integral approach seems to be more effective, at least from perturbative point of view.
\section{Conclusions}
Caldeira-Leggett model of reservoir is generalized to a reservoir modeled by a continuum of real Klein-Gordon fields instead of harmonic oscillators. A quantum Langevin type dissipative equation is obtained for the scalar field. The susceptibility of the medium is defined in terms of the reservoir Green's function and the coupling function satisfying causality condition. The connection between the coupling function and the susceptibility of the medium is found to be a Hankel transform from which the coupling function can be determined in terms of the susceptibility of the medium. Noise currents and their fluctuation-dissipation relation are obtained. In a homogeneous medium or reservoir, explicit form of the quantum scalar field, and its large-time limit, are found. Knowing the explicit form of the quantum field, one can proceed and construct Fock's space or find important quantities like the two-point function or Feynman Green's function. Based on the structure of the article, the generalization to an interacting dissipative field theory is straightforward in the framework of path integrals, at least perturbatively.


\begin{thebibliography}{22}
\bibitem{Casimir} D.A.R. Dalvit, P.W. Milonni, D.C. Roberts
and F.S.S. Rosa eds., Lecture Notes in Physics 834
(Springer-Verlag, 2011) p.345.
\bibitem{Guth} A. H. Guth, Phys. Rev D\textbf{23}, 347 (1981).
\bibitem{Brand} Brandenberger, Rev. of Mod. Phys. 57, 1 (1985).
\bibitem{Huttner} B. Huttner and S. M. Barnett, Phys.Rev. A\textbf{46}, 4306 (1992).
\bibitem{Kheir} F. Kheirandish and M. Soltani, Phys.Rev. A\textbf{78}, 012102 (2008); F. Kheirandish and M. Jaffari, Phys.Rev. A\textbf{86}, 022503 (2012); 
F. Kheirandish, M. Soltani, and M. Jafari, Phys.Rev. A\textbf{84}, 062120 (2011).
\bibitem{Amoo} M. Amooshahi, Eur. Phys. J. D\textbf{54}, 115 (2009).
\bibitem{Caldeira} A. O. Caldeira and A. J. Leggett, Phys. Rev. Lett. 46, 211 (1981); Ann. Phys. 149, 374
(1983); Physica 121A, 587 (1983).
\bibitem{Weiss} U. Weiss, Quantum Dissipative Systems (1993, World Scientific) and the references
therein.
\bibitem{Voss} R. F. Voss and R. A. Webb, Phys. Rev. Lett. 47, 265 (1981).
\bibitem{Martinis} J. M. Martinis, M. H. Devoret and J. Clarke, Phys. Rev. Lett. 55, 1543 (1985).
\bibitem{Gradshteyn} I. S. Gradshteyn and I. M. Ryzhik, Table of Integrals, Series,
and Products (Academic, New York, 2007).
\end{thebibliography}
\end{document}